\documentclass[aps,prl,twocolumn]{revtex4}
\usepackage{color, graphicx}%
\usepackage{dcolumn}
\usepackage{amsmath}
\usepackage{amsfonts}
\usepackage{latexsym}
\newcommand{\ve}{\varepsilon}

\newcommand{\tm}{\tilde m}

\newcommand{\si}{S}

\newcommand{\bx}{\mathbf{x}}
\newcommand{\bk}{\mathbf{k}}

\newcommand{\cB}{{\cal{B}}}

  \def\lsim{\mathrel {\vcenter {\baselineskip 0pt \kern 0pt
    \hbox{$<$} \kern 0pt \hbox{$\sim$} }}}
    \def\gsim{\mathrel {\vcenter {\baselineskip 0pt \kern 0pt
    \hbox{$>$} \kern 0pt \hbox{$\sim$} }}}

\begin{document}
\bibliographystyle{apsrev}
\title{Spin current as a response to external stress}

\author{Prashant Sharma}
\email[]{E-mail: psharma@anl.gov}
%\thanks{}

\affiliation{Materials Science Division, Argonne National Laboratory, Argonne IL 60439
}

\date{\today}

\begin{abstract}
\noindent 
It is theoretically predicted that a traveling shear wave will create a spin current in certain direct-gap (for example III-V compound) semiconductors with contributions from both the valence bands and the conduction band (for $n$-doped semiconductors). We show that this spin-current is a property of the Fermi-Dirac sea, and is controlled by a geometric phase accumulated by the strain-induced Rashba parameters in a cycle.   
\end{abstract}

\maketitle
%%%%%%%%%
%%%%%%%%%%%%%%%%%%%%%%%%%%%%%%%
%{\it{Introduction:}}\\
%%%
 In recent years there has been a great deal of experimental and theoretical interest in the manipulation of electronic spin in semiconducting materials~\cite{zutic-04}. Generating spin current in a controlled way has emerged as one of the central problems of spintronics~\cite{wolf-science,sharma-05}. Early interest in the obtaining spin polarized current by adiabatic pumping through quantum dots~\cite{Marcus-03} was not without experimental difficulties~\cite{brouwer-01}. More recently, progress has been made in generating spin currents of fixed polarization direction in bulk semiconductors~\cite{kato-2004} using the extrinsic spin-Hall effect~\cite{dyakonov-1971, hirsch-1999,engel-2005}, and in two-dimensional layers of $p$-GaAs~\cite{wunderlich-2005} using the intrinsic spin-Hall effect~\cite{murakami-03,bernevig-2005,sinova-04}. 

In contrast to charge current, spin angular momentum current can occur without net transport of particles. Therefore one expects to find a method for creating a spin current \emph{without} using external electric fields which lead to Fermi surface effects. In this paper we show that it is possible to use the coupling of electrons to local deformations in the crystal in such a way as to transfer external angular momentum to the electron's spin~\footnote{Transferring spin angular momentum to the lattice can be useful for spin detection and manipulation (see Ref.~\cite{mohanty-fulde-04}).}. This offers an alternative to the aforementioned methods for generating and controlling spin currents. 

Under shear stress that breaks crystalline symmetry locally, the motion of a Bloch electron is affected depending on its spin state. In a semiclassical picture, two different effects occur together: (i) Because of local lattice deformations that break inversion symmetry, if a particle moving in (for example) the positive $z$ direction enters such a region and acquires a deflection in the positive $x$-direction (dashed line in Fig.~1.), then the oppositely moving particle (solid line in Fig.~1) is deflected with a different (in magnitude) $x$ component of the velocity. (ii) Because of strain-induced spin-orbit interaction the quasiparticle's energy in this new environment depends on its spin direction and is odd in its velocity. Its effect is, therefore, the same as a spin-dependent vector potential for the quasiparticle. If the strain is position dependent along the $x$-axis (for example) so that the vector potential has a non-zero curl, then the resulting orbital magnetic field is spin-dependent, but insensitive to the direction of velocity. Let us say the situation is such that the magnetic field has only a $y$-axis component. The Lorentz force that acts on quasiparticles moving in the $z$-direction (see Fig.~1.) is such as to deflect the up-spins (blue dashed line) in the positive $x$-direction and the down-spins (red dashed line) to the negative $x$-direction. Under a reversal of velocities (solid lines in the figure) an opposite deflection will occur. Because of the asymmetrical scattering (i), however, these two events will not cancel each other. Therefore, as this lattice distortion moves through the crystal it will create a net spin angular momentum current flowing in the $x$-direction.
\begin{figure}[ht]
%  \begin{center}
   \includegraphics[scale=.25]{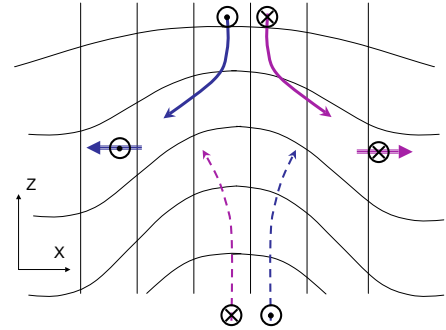}
   \caption{Schematic of time reversed semiclassical trajectories of spin up and down particles. The presence of local lattice deformation and the induced spin-orbit interaction breaks time-reversal symmetry and creates unequal deflection of these trajectories leading to a net spin current.}
%  \end{center}
\end{figure}

%%%%%%%%
%{\it Hamiltonian and the spin current}:\\
%
To put the above arguments on a firm footing, let us consider a direct gap semiconductor crystal with cubic and time-reversal symmetry. The Hamiltonian for conduction band electrons conserves the spin $\vec S$ and is written in the spherical approximation as
\begin{subequations}\label{htotal}
\begin{align}
H_{0c}=\frac{\hbar^2\bk^2}{2m_{\rm c}},
\end{align} 
where $m_{\rm c}$ denotes the effective mass in the conduction band and $\bk$ is its crystal momentum. The Hamiltonian for the valence band is more involved if the bands are degenerate. For definiteness, we will consider the case, where at the zone center (the $\Gamma$ point) the valence band is triply degenerate (without spin) and well-separated from the conduction band. The valence band Hamiltonian is therefore described by the orbital angular momentum projection $l=\pm 1,0$ with a (negative) reduced effective mass for the electron. Because of spin-orbit coupling the conserved quantity is the total angular momentum $\vec J=\vec L+\vec S$, and the helicity $\bk\cdot\vec S/k$. Therefore the orbital angular momentum bands are further split into the (doubly degenerate) heavy hole (HH) and light hole (LH) bands with $J=3/2$ and helicity quantum numbers $\pm 3/2$ and $\pm1/2$, and the $J=1/2$ and helicity $\pm1/2$ SH-band. The Hamiltonian near the $\bk=0$ ($\Gamma$) point is given by the Luttinger Hamiltonian~\cite{luttinger-56}:
\begin{align}\label{hlut}
H_0=\frac{\hbar^2}{2m}(\gamma_1+\frac{5}{2}\gamma)\bk^2-\frac{\gamma}{m}(\bk\cdot\vec S)^2
\end{align}
where $m$ is the free electron mass and the $\gamma$s are Luttinger's parameters, and the spin $\vec S$ is in the spin-3/2 representation of SU($2$). In the spherical approximation the energy spectrum of the various bands is obtained as:
\begin{align}
H_{0\mu}=\frac{\hbar^2\bk^2}{2m_\mu},\,\,{\mu={\rm HH},\,{\rm LH},\,{\rm SH}},\,{\rm c},
\end{align}
where $m_\mu$ denotes the effective mass of the electron in the $\mu$-th band. Let us note here that the spherical approximation comes with an upper cut-off that we denote by $\bk_c$.

When the crystal is subject to a transverse (shear) wave, both inversion symmetry as well as time-reversal symmetry (because of the traveling wave-form) are broken. To consider transport along one direction we choose an acoustic wave traveling along the $x$-axis with particle motion along the $y$- and $z$-axes so as to create time-dependent strains $\ve_{xy}\equiv\phi_z(x,t)$ (along the $z$ crystalline axis), $\ve_{yz}\equiv\phi_{x}(x,t)$ (along the $x$-crystalline axis), and $\ve_{xz}\equiv\phi_y(x,t)$ (along the $y$ crystalline axis). To describe the response of the electrons to this perturbation we adopt the semiclassical approach of Ref.~\cite{niu-sundaram-99} in which we consider a wave-packet centered at the coordinate $\bx\equiv(x,y,z)$ at a given time with its spread small compared to the wavelength $2\pi/q$ of the acoustic perturbation. The local Hamiltonian for such a wave-packet can be written down in the basis state of the Bloch wave $|S\mu\bk,\bx,t\rangle$. In the ``clean" limit---when the wave-vector of the sound wave $q>1/l_p$, the inverse particle scattering length---the electrons move in phase with the traveling shear wave. The dominant contribution to spin-orbit interaction from such a strain is linear in the crystal momentum $\bk$ and affects both the conduction and valence bands~\cite{pikus-84} so that the energy of the wave-packet is
\begin{align}
H^{\ve}_\mu(t)&={C_\mu}\Big[\phi_x(x,\!t)\Big(\si^y k_z-\si^z k_y\Big)
%\nonumber\\
+\phi_y(x,\!t)\Big(\si^z k_x
\nonumber\\
&-\si^x k_z\Big)
+\phi_z(x,\!t)\Big(\si^x k_y-\si^y k_x\Big)\Big],\label{het}
\end{align}
where the spin of the quasiparticle is $\vec S$, and the velocity $C_\mu$ depends on the semiconductor band structure~\cite{pikus-84}.
Besides this Rashba correction, $H^{\ve}_\mu$, the strain also creates a deformation potential~\cite{bir-pikus72,luthi05,khan84,niu-sundaram-99}:
\begin{align}
H^{{\rm def}}_{\mu}(t)&=-\phi_x(x,t)\frac{\hbar^2k_zk_y}{\tm_\mu}-\phi_z(x,t)\frac{\hbar^2k_xk_y}{\tm_\mu}
\nonumber\\
        &~~~~~-\phi_y(x,t)\frac{\hbar^2k_xk_z}{\tm_\mu},
\end{align}
where $\tm_\mu^{-1}=m(m^{-1}_\mu-m^{-1})^2$, and we have written the electron-phonon deformation potential in the deformable ion approximation~\footnote{In the alternative rigid ion approximation (see for example Ref.~\cite{khan84}), the deformation potential for acoustic phonons will acquire an additional term $\sum_{a,b}V_{ab}(\bx)\ve_{ab}(\bx,t)$. In the deformable ion approximation this term is zero~\cite{bir-pikus72}. Including this term does not bring anything new because the strains we consider are purely shear.}. 
With these approximations we obtain the energy of the electrons in the $\mu$-th band perturbed by the acoustic wave
\begin{align}
H_\mu=H_{0\mu}+H^{{\rm def}}_\mu(t)+H^{\ve}_\mu(t),\,\,{\mu={\rm c, HH, LH, SH}}
\end{align}
\end{subequations}
The time-dependent perturbations $H^{{\rm def}}_\mu(t)+H^{\ve}_\mu(t)$ change the crystal momentum $k_x$ according to the semiclassical Bloch equations of motion:
\begin{subequations}\label{kxoft}
\begin{align}
\frac{d k_x}{dt}&= \hat E(t) + \hat F(t)\;k_x\\
\hat E(t)&=-\hbar\frac{\partial\phi_x}{\partial x}\frac{k_zk_y}{\tm_\mu}+\frac{C_\mu}{\hbar}\left[\frac{\partial\phi_z}{\partial x}k_y-\frac{\partial\phi_y}{\partial x}k_z\right]\si^x
\nonumber\\
&+\frac{C_\mu}{\hbar}\frac{\partial\phi_x}{\partial x}\left[\si^yk_z-\si^zk_y\right]
,\\
\hat F(t)&=-\frac{\hbar k_y}{\tm_\mu}\frac{\partial\phi_z}{\partial x}-\frac{\hbar k_z}{\tm_\mu}\frac{\partial\phi_y}{\partial x}
-\frac{C_\mu}{\hbar}\frac{\partial\phi_z}{\partial x}\si^y
+\frac{C_\mu}{\hbar}\frac{\partial\phi_y}{\partial x}\si^z.
\end{align}
\end{subequations}
Note that the wave packet's velocity transverse to the wave propagation direction also changes:
\begin{subequations}\label{v-transverse}
\begin{align}
\frac{dy}{dt}&=\frac{\hbar k_y}{m_\mu}+\frac{C_\mu}{\hbar}\left(\vec\phi\times\vec\si\right)_y
+\phi_x\frac{\hbar k_z}{\tm_\mu}+\phi_z\frac{\hbar k_x}{\tm_\mu}\\
\frac{dz}{dt}&=\frac{\hbar k_z}{m_\mu}+\frac{C_\mu}{\hbar}\left(\vec\phi\times\vec\si\right)_z
+\phi_x\frac{\hbar k_y}{\tm_\mu}+\phi_y\frac{\hbar k_x}{\tm_\mu}.
\end{align}
\end{subequations}
Eqn.~\ref{kxoft} can be readily integrated to give
%\label{kxt}
\begin{align*}
k_x(t)=\int_0^t \frac{dt'}{2}\!\!\left\{e^{\int_{t'}^tdt''\hat F(t'')}\,,\,\hat E(t')\right\} + e^{\int_{0}^tdt''\hat F(t'')}\;k_{x0},
\end{align*}
where $k_{x0}$ denotes the crystal momentum in the bare (unperturbed) crystal which is the situation for time $t\leq 0$, and $\{\ A,B\}$ denotes the symmetrized product.
 
 The current response to this time-dependent perturbation can be calculated analytically if we restrict ourselves to terms up to $\phi_a^2\ll 1$. This scheme requires that the strains satisfy $(v_c/v_s)|\phi_a|\ll 1$, where $v_c={\rm max}[C_\mu,\hbar k_c/\tm_\mu]$, $k_c$ the upper cut-off, and $v_s$ the velocity of sound. Upon expanding the exponential in the equation above we determine $k_x(t)$ as a function of the spin $\vec S$ and the equilibrium momenta $k_{x0},k_{y,z}$:
% 
%\begin{subequations}
\begin{align}\label{kxt-2}
k_x&\approx k_{x0}\left[1+\int_{0}^t\!\!dt_1\hat F(t_1)-\int_{0}^t\!\!\frac{dt_1}{2}\frac{dt_2}{2}
\{\hat F(t_1),\hat F(t_2)\}\right]
\\\nonumber
&+\int_0^tdt'\hat E(t') + \int_0^t \frac{dt'}{2}\int_{t'}^t{dt''}\{\hat F(t''),\hat E(t')\}
\end{align}
%\end{subequations}
% 
The spin $\vec S$ also acquires {\it additional} dynamics from the perturbation. This is identified as arising from an effective Zeeman field $\vec\cB$ with components (in units of $C_\mu$)
%
%\begin{subequations}
%
\begin{align*}
\cB_x=\phi_yk_z-\phi_zk_y;\,
\cB_y=\phi_z k_x-\phi_x k_z;\,
\cB_z=\phi_x k_y-\phi_y k_x.
\end{align*}
%\end{subequations}
%
The perturbation gives rise to a (time-dependent) Rashba splitting between the $S$ bands 
$
2|S|C_\mu\cB=2|S| C_\mu\sqrt{\cB_x^2+\cB_y^2+\cB_z^2}.
$ 
and the {\it change} in the equations of motion 
of the spin because of the Hamiltonian (Eqn.~\ref{het}) is obtained using the SU($2$) algebra $[S^a,S^b]=i\hbar\epsilon_{abc}S^c$:
\begin{align}\label{dsdot}
\frac{d\vec\si}{dt}&=C_\mu\vec\si\times\vec{\cal B}
\end{align}
The generalized current density in the $a$-direction of charge ($b=0$), or the $b$-th component of spin $\vec S$ is given by
\begin{align}
j_a^b(t)&=\frac{1}{\Omega}\sum_{S,\mu,\bk_0}\frac{n_\mu(\bk_0)}{2}\langle{S\mu}\bk,\bx,t\vert \{\frac{\partial H_{0\mu}}{\hbar\partial k_a},\si^b(t)\}\vert{S\mu}\bk,\bx,t\rangle,
\end{align}
where $\Omega$ is the unit cell volume, $S^{b=0}\equiv e$, the electron charge, and 
$n_\mu(\bk_0)$ denotes the quasi-particle occupation in the equilibrium state with momentum $\bk_0$, band index $\mu$, and with quantum number $S$, and $|S\mu\bk,\bx,t\rangle$ denotes the Bloch wave at coordinate $\bx$ and the time $t$. We have not included the $S$- index for the Fermi occupation number since in equilibrium these states are degenerate and are equally occupied.  Because the momenta $k_y$ and $k_z$ are conserved there are no charge currents along the $y$- and $z$-directions. The only possibility of a non-zero charge current is along the $x$-direction because $k_x$ is not conserved. However, the change in $k_x$ (see Eq.~\ref{kxt-2}) does not create any charge current as $\hat E$ and the explicitly spin-dependent part of $\hat F$ sum to zero over the $S$-bands, while the product of $k_{x0}$ and the spin-independent part of $\hat F$ averages to zero over the zone. Furthermore, there is no spin current along the $y$- and $z$-directions because the momenta $k_{y,z}$ are uncorrelated with spin. We therefore focus on the spin current
\begin{subequations}
\begin{align}
j_x^b(t)=\sum_{\mu,\bk_0,S}\frac{n_\mu(\bk_0)}{2\Omega }\langle{S\mu}\bk,\bx,t\vert\{\frac{\hbar k_x(t)}{m_\mu},S^b(t)\}\vert{S\mu}\bk,\bx,t\rangle,
\end{align}
that can be calculated using Eq.~\ref{kxt-2}. Only those terms in the summand contribute that are: (i) functions of $k^2_{x0},\,k_y^2,\,k_z^2$, and (ii) quadratic in the spin $\vec S$. Therefore we look at those terms in (\ref{kxt-2}) that are functions of $k^2_{a0}$ and are linear in $\vec S$. Only the last term in Eq.~\ref{kxt-2} contributes, and we focus now on simplifying it. In doing so we shall make use of the smallness of the strain parameters 
$|\phi_a|\ll 1$: Because $\{\hat E(t'),\hat F(t'')\}$ is already second order in $\phi_a$ we will neglect the {\it additional} dynamics of the spin (Eqn.~\ref{dsdot}) in the evaluation of $j_x^b$ as its time evolution will bring in higher powers of $\phi_a$. We thus obtain
\begin{align}\label{jx-s-spin}
j_x^b(t)&=\frac{\hbar}{\Omega}\sum_{\bk_0,\mu,S}\frac{n_\mu(\bk_0)}{2m_\mu}
\int_0^tdt'\frac{C_\mu}{v^2_s\tm_\mu}\Big\langle\Big\{\si^b(t),
\nonumber\\
&S^x(t)[\Delta\phi_y\dot\phi_y(t')k^2_z
-\Delta\phi_z\dot\phi_z(t')k^2_y]
\nonumber\\
&-S^y(t)\Delta\phi_y\dot\phi_x(t')k^2_z+S^z(t')\Delta\phi_z\dot\phi_x(t')k^2_y\Big\}\Big\rangle
\end{align} 
\end{subequations}
where $\Delta\phi_a\equiv\phi_a(t)-\phi_a(t')$. The terms we have neglected involve time derivatives of the spin operator. For conduction band spin this of course vanishes. For the valence band, however, (\ref{dsdot}) is not the only dynamics of the spin, as only helicity is a good quantum number. The form of the Luttinger Hamiltonian (\ref{hlut}) implies that the time derivative of spin will be quadratic in the spin. Together with $S^b(t)$ present in (\ref{jx-s-spin}) this will bring in a three spin expectation value in the equilibrium state. This vanishes upon doing the helicity sum, thus justifying the neglect of spin dynamics in determining the spin current.  

Let us write $\phi_{y,z}(x,t)\equiv f_{y,z}\phi_x(x,t)+g_{y,z}\tilde\phi_x(x,t)$, where $\tilde\phi_x$ is out of phase with $\phi_x$, and $f_{y,z},\, g_{y,z}$ are time independent.
Using this we obtain the {\it dc} spin current:
%
%\begin{subequations}
%
\begin{align}
j_x^x=&
\frac{\hbar}{\Omega}\sum_{\bk_0,\mu}\frac{n_\mu(\bk_0)}{2}\frac{C_\mu}{v_s^2\hbar^2}\frac{\hbar^2k_z^2}{\tm_\mu m_\mu}{\rm tr}({S^x}^2)\left(\overline{\phi_y^2}-\overline{\phi_z^2}\right)
\\
j_x^{y(z)}&=\frac{\hbar}{\Omega}\sum_{\bk_0,\mu}\frac{n_\mu(\bk_0)}{2}\frac{C_\mu}{v_s^2\hbar^2}\frac{\hbar^2k_z^2}{\tm_\mu m_\mu}{\rm tr}({S^{y(z)}}^2)\left(\pm f_{y(z)}\right)\overline{\phi_x^2},\nonumber
\end{align}
%
%\end{subequations}
%
where the horizontal bar denotes time average over the wave's period. We immediately see that the 
{\it dc} spin current is determined by the angular momentum imparted to the quasi-particles along the polarization axis by the traveling wave. Within the summand, the quantity $\tm^{-1}_\mu\equiv m^{-1}-2m^{-1}_\mu$. Over most of the zone the semiconductor band mass is much smaller than the electron mass and we can use $\tm^{-1}_\mu\approx-2m^{-1}_\mu$. It follows that the net spin current is {\it independent} of the sign of the effective mass: All bands contribute together. 

The spin current is controlled by the geometric phase that a quasiparticle accumulates (in a period) as it moves in a trajectory tracking the $\vec\phi$-parameter. The Rashba Hamiltonian (\ref{het}) implies a spin-dependent (non-abelian) vector potential for the $\mu$-th band: 
$\vec A=m_\mu (C_\mu/\hbar)\vec\phi\times\vec\si$. The corresponding magnetic field is
\begin{align}\label{orbital-b}
B_x&=0,\,B_y=m_\mu(C_\mu/\hbar)[\partial_x\phi_y\si^x-\partial_x\phi_x\si^y],
\nonumber\\
B_z&=m_\mu(C_\mu/\hbar)[\partial_x\phi_z\si^x-\partial_x\phi_x\si^z].
\end{align}
The motion of the particle in the $x-z$ plane is controlled by $\phi_y$ so that the  
flux of this magnetic field through a unit area oriented in the $\hat y$ direction is proportional 
to $B_y\phi_y$. Similarily, the motion of the particle in the $x-y$ plane is controlled by $\phi_z$ and the corresponding flux is proportional to $B_z\phi_z$. Using (\ref{orbital-b}) and comparing the resulting expression with the spin current (\ref{jx-s-spin}) it is seen that the difference of these two fluxes determines the spin current that is proportional to ${\rm tr}[S^b(B_y\phi_y-B_z\phi_z)]$.

 The magnitude of the spin current can be estimated as follows. Consider the case when $\overline{\phi_y^2}=\overline{\phi_z^2}$ and let us also choose $f_y=0=g_z$ and $f_z=g_y\neq0$, so that the dc current is only polarized in the $z$-direction:
\begin{align}\label{jxz-dc}
j_x^z=-\frac{\overline{\phi_x^2}}{\Omega}
\sum_{\bk_0,\mu}\frac{C_\mu}{v_s^2}
\frac{n_\mu(\bk_0)}{2}
\frac{\hbar^2k_z^2}{\tm_\mu m_\mu}
{\rm tr}({S^{z}}^2)/\hbar
\end{align}
Such a choice of shear strains can be obtained by applying shear stresses schematically drawn in Fig.~2. The contribution of the integral in (\ref{jxz-dc}) comes from parts of the zone where the effective mass is significantly different from the electron mass. To estimate a lower bound for the current let us limit the summation range to an upper cut-off wavevector $k_c$, alluded to earlier, and consider $n$-doped semiconductor at low temperature. The lower bound is estimated considering only the conduction band ($\mu={\rm c}$) for $n$-doped semiconductor:
\begin{align}
j_x^z\gsim\frac{\hbar}{5\pi^2}\frac{\overline{\phi_x^2}}{v_s^2}\frac{\hbar^2k_c^2}{m_\mu}\frac{C_\mu k_c^3}{m_\mu}\frac{{\rm tr}({S^{z}}^2)}{\hbar^2},
\end{align}
using GaAs data: $C_{\rm c}=8\times 10^7{\rm cm/s}$~\cite{pikus-84}, mass ratio $m_{\rm c}/m=0.05$ that is fairly representative of most semiconductors, and taking $k_c=3\times10^6{\rm cm}^{-1}=10^{-4}\lambda_c^{-1}$ ($k_c=k_F$ for low doping with $E_F=3\;{\rm meV}$, $\lambda_c$ is the electron's Compton wavelength), and the velocity of sound $v_s=10^5{\rm cm/s}$. The value of strain $(\overline{\phi_x^2})^{1/2}$ will depend on how the acoustic wave is generated. 
One way to generate such a wave is to contact the crystal by a piezoelectric material in which acoustic waves are induced by applied ac electric field. This conventional technique gives (for electric field strength of $10{\rm V/mm}$ and using GaAs elastic and piezoelectric constants) a strain 
$(\overline{\phi^2})^{1/2}=10^{-6}$~\cite{luthi05}. Furthermore, we are assuming that the shear wave has a wavelength comparable to, or smaller than the momentum scattering length $l_p$. At low (liquid nitrogen) temperatures, the value for $l_p\approx1{\rm \mu m}$ in semiconductors at low doping~\cite{pikus-84}. Acoustic waves of such wavelengths have been obtained in semiconductor crystals~\cite{mohanty-05}. We also note that the spin lifetime $\tau_s\lsim{\rm ns}$~\cite{pikus-84,kato-2004} in several materials, which allows experiments at ${\rm GHz}$ frequency. Using these parameters we obtain 
$(e/\hbar) j_x^z\gsim 10^2\;{\rm Amp/cm^2}$, which is not unreasonable, especially since a spin current smaller than this magnitude has been detected~\cite{kato-2004}, and also because the contribution from the valence band comes from the entire zone and is therefore much larger than this lower bound. 
\begin{figure}[b]\label{fig-jxz}
%  \begin{center}
   \includegraphics[scale=.35]{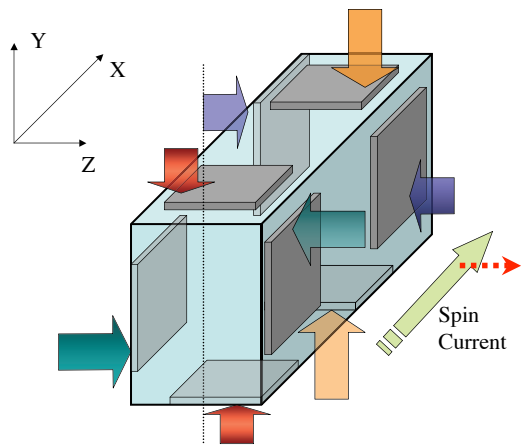}
   \caption{Schematic of a cubic crystal subject to shear stresses about the $x,\,y$ and $z$ axes. As these stresses are applied periodically in time and out of phase (depicted by different colors and sizes) along the $x$-axis they generate a traveling wave that creates a spin current polarized along the $z$-axis. The direction of polarization can be changed by altering the phase relationships between the stresses.}
%  \end{center}
\end{figure}
%

%%%%%%%%%%%%%%%%%%%%%%%%
%\clearpage
%%%%%%%%%%%%%%%%%%%%%%%
%{\it Conclusion}:\\
Our calculation above may seem to be limited by the choice of an unperturbed Hamiltonian that has spherical dispersion in the valence band. It should be emphasized, therefore, that  the only feature of the valence band that is essential is that, besides crystal momentum, the electron's total angular momentum $\vec J$ is conserved. The perturbation alters the crystal momentum along one ($x$) direction thereby leaving only $\vec J$ in that direction conserved. Thus $dJ_x/dt=0=\hbar(\vec v\times\vec k)_x+d S_x/dt$. Using Eqns.~\ref{v-transverse} and ~\ref{kxt-2}, the first term, which is proportional to the spin current, can be calculated and its projection onto $S^b(t)$ gives Eqn.~\ref{jx-s-spin} for spin current.  

 In conclusion let us remark that unlike the intrinsic spin-Hall effect~\cite{murakami-03,sinova-04,bernevig-2005}, the spin-acoustic effect described here arises from explicitly breaking time-reversal symmetry. This leads to contributions from all occupied bands that are affected by strain-induced spin-orbit interactions. In this sense the spin current can be thought of as a diamagnetic response of the semiconductor to effective spin-dependent magnetic fields created by non-uniform stress.

 The author acknowledges support from the U.S. Dept. of Energy, under Contract No. W-31-109-ENG-38, and thanks P.~W.~ Brouwer, C.~L.~Henley, K.~Matveev and C.~Chamon for useful comments on the manuscript.

%%
%\bibliography{spintronics_bib}

\begin{thebibliography}{21}
\expandafter\ifx\csname natexlab\endcsname\relax\def\natexlab#1{#1}\fi
\expandafter\ifx\csname bibnamefont\endcsname\relax
  \def\bibnamefont#1{#1}\fi
\expandafter\ifx\csname bibfnamefont\endcsname\relax
  \def\bibfnamefont#1{#1}\fi
\expandafter\ifx\csname citenamefont\endcsname\relax
  \def\citenamefont#1{#1}\fi
\expandafter\ifx\csname url\endcsname\relax
  \def\url#1{\texttt{#1}}\fi
\expandafter\ifx\csname urlprefix\endcsname\relax\def\urlprefix{URL }\fi
\providecommand{\bibinfo}[2]{#2}
\providecommand{\eprint}[2][]{\url{#2}}

\bibitem[{\citenamefont{Zutic et~al.}(2004)\citenamefont{Zutic, Fabian, and
  Sarma}}]{zutic-04}
\bibinfo{author}{\bibfnamefont{I.}~\bibnamefont{Zutic}},
  \bibinfo{author}{\bibfnamefont{J.}~\bibnamefont{Fabian}}, \bibnamefont{and}
  \bibinfo{author}{\bibfnamefont{S.~D.} \bibnamefont{Sarma}},
  \bibinfo{journal}{Rev. Mod. Phys.} \textbf{\bibinfo{volume}{76}},
  \bibinfo{pages}{323} (\bibinfo{year}{2004}).

\bibitem[{\citenamefont{{Wolf {\it et al.}}}(2001)}]{wolf-science}
\bibinfo{author}{\bibfnamefont{S.~A.} \bibnamefont{{Wolf {\it et al.}}}},
  \bibinfo{journal}{Science} \textbf{\bibinfo{volume}{294}},
  \bibinfo{pages}{1488} (\bibinfo{year}{2001}).

\bibitem[{\citenamefont{Sharma}((2005) {\rm and references
  therein.})}]{sharma-05}
\bibinfo{author}{\bibfnamefont{P.}~\bibnamefont{Sharma}},
  \bibinfo{journal}{Science} \textbf{\bibinfo{volume}{307}},
  \bibinfo{pages}{531} (\bibinfo{year}{(2005) {\rm and references therein.}}).

\bibitem[{\citenamefont{{Watson, {\it et al.}}}(2003)}]{Marcus-03}
\bibinfo{author}{\bibfnamefont{S.~K.} \bibnamefont{{Watson, {\it et al.}}}},
  \bibinfo{journal}{Phys. Rev. Lett.} \textbf{\bibinfo{volume}{91}},
  \bibinfo{pages}{258301} (\bibinfo{year}{2003}).

\bibitem[{\citenamefont{Brouwer}(2001)}]{brouwer-01}
\bibinfo{author}{\bibfnamefont{P.~W.} \bibnamefont{Brouwer}},
  \bibinfo{journal}{Phys. Rev. B} \textbf{\bibinfo{volume}{63}},
  \bibinfo{pages}{121303} (\bibinfo{year}{2001}).

\bibitem[{\citenamefont{Kato et~al.}(2004)\citenamefont{Kato, Myers, Gossard,
  and Awschalom}}]{kato-2004}
\bibinfo{author}{\bibfnamefont{Y.~K.} \bibnamefont{Kato}},
  \bibinfo{author}{\bibfnamefont{R.~C.} \bibnamefont{Myers}},
  \bibinfo{author}{\bibfnamefont{A.~C.} \bibnamefont{Gossard}},
  \bibnamefont{and} \bibinfo{author}{\bibfnamefont{D.~D.}
  \bibnamefont{Awschalom}}, \bibinfo{journal}{Science}
  \textbf{\bibinfo{volume}{306}}, \bibinfo{pages}{1910} (\bibinfo{year}{2004}).

\bibitem[{\citenamefont{D'yakonov and Perel}(1971)}]{dyakonov-1971}
\bibinfo{author}{\bibfnamefont{M.~I.} \bibnamefont{D'yakonov}}
  \bibnamefont{and} \bibinfo{author}{\bibfnamefont{V.~I.} \bibnamefont{Perel}},
  \bibinfo{journal}{Phys. Lett. A} \textbf{\bibinfo{volume}{35}},
  \bibinfo{pages}{459} (\bibinfo{year}{1971}).

\bibitem[{\citenamefont{Hirsch}(1999)}]{hirsch-1999}
\bibinfo{author}{\bibfnamefont{J.~E.} \bibnamefont{Hirsch}},
  \bibinfo{journal}{Phys. Rev. Lett.} \textbf{\bibinfo{volume}{83}},
  \bibinfo{pages}{1834} (\bibinfo{year}{1999}).

\bibitem[{\citenamefont{Engel et~al.}(2005)\citenamefont{Engel, Halperin, and
  Rashba}}]{engel-2005}
\bibinfo{author}{\bibfnamefont{H.-A.} \bibnamefont{Engel}},
  \bibinfo{author}{\bibfnamefont{B.~I.} \bibnamefont{Halperin}},
  \bibnamefont{and} \bibinfo{author}{\bibfnamefont{E.~I.}
  \bibnamefont{Rashba}}, \bibinfo{journal}{Phys. Rev. Lett.}
  \textbf{\bibinfo{volume}{95}}, \bibinfo{pages}{166605}
  (\bibinfo{year}{2005}).

\bibitem[{\citenamefont{Wunderlich et~al.}(2005)\citenamefont{Wunderlich,
  Kaestner, Sinova, and Jungwirth}}]{wunderlich-2005}
\bibinfo{author}{\bibfnamefont{J.}~\bibnamefont{Wunderlich}},
  \bibinfo{author}{\bibfnamefont{B.}~\bibnamefont{Kaestner}},
  \bibinfo{author}{\bibfnamefont{J.}~\bibnamefont{Sinova}}, \bibnamefont{and}
  \bibinfo{author}{\bibfnamefont{T.}~\bibnamefont{Jungwirth}},
  \bibinfo{journal}{Phys. Rev. Lett.} \textbf{\bibinfo{volume}{94}},
  \bibinfo{pages}{047204} (\bibinfo{year}{2005}).

\bibitem[{\citenamefont{Murakami et~al.}(2003)\citenamefont{Murakami, Nagaosa,
  and Zhang}}]{murakami-03}
\bibinfo{author}{\bibfnamefont{S.}~\bibnamefont{Murakami}},
  \bibinfo{author}{\bibfnamefont{N.}~\bibnamefont{Nagaosa}}, \bibnamefont{and}
  \bibinfo{author}{\bibfnamefont{S.-C.} \bibnamefont{Zhang}},
  \bibinfo{journal}{Science} \textbf{\bibinfo{volume}{301}},
  \bibinfo{pages}{1348} (\bibinfo{year}{2003}).

\bibitem[{\citenamefont{Bernevig and Zhang}(2005)}]{bernevig-2005}
\bibinfo{author}{\bibfnamefont{B.~A.} \bibnamefont{Bernevig}} \bibnamefont{and}
  \bibinfo{author}{\bibfnamefont{S.-C.} \bibnamefont{Zhang}},
  \bibinfo{journal}{Phys. Rev. Lett.} \textbf{\bibinfo{volume}{95}},
  \bibinfo{pages}{16801} (\bibinfo{year}{2005}).

\bibitem[{\citenamefont{{Sinova, {\it et al.}}}(2004)}]{sinova-04}
\bibinfo{author}{\bibfnamefont{J.}~\bibnamefont{{Sinova, {\it et al.}}}},
  \bibinfo{journal}{Phys. Rev. Lett.} \textbf{\bibinfo{volume}{92}},
  \bibinfo{pages}{126603} (\bibinfo{year}{2004}).

\bibitem[{\citenamefont{Luttinger}(1956)}]{luttinger-56}
\bibinfo{author}{\bibfnamefont{J.~M.} \bibnamefont{Luttinger}},
  \bibinfo{journal}{Phys. Rev.} \textbf{\bibinfo{volume}{102}},
  \bibinfo{pages}{1030} (\bibinfo{year}{1956}).

\bibitem[{\citenamefont{Meier and {Zakharchenya, eds.}}(1984)}]{pikus-84}
\bibinfo{author}{\bibfnamefont{F.}~\bibnamefont{Meier}} \bibnamefont{and}
  \bibinfo{author}{\bibfnamefont{B.~P.} \bibnamefont{{Zakharchenya, eds.}}},
  \emph{\bibinfo{title}{Optical Orientation}} (\bibinfo{publisher}{North
  Holland}, \bibinfo{year}{1984}), chap. \bibinfo{chapter}{2 {by G. E. Pikus
  and A. N. Titkov}}.

\bibitem[{\citenamefont{Bir and Pikus}(1974)}]{bir-pikus72}
\bibinfo{author}{\bibfnamefont{G.~L.} \bibnamefont{Bir}} \bibnamefont{and}
  \bibinfo{author}{\bibfnamefont{G.~E.} \bibnamefont{Pikus}},
  \emph{\bibinfo{title}{Symmetry and strain-induced effects in semiconductors}}
  (\bibinfo{publisher}{New York, Wiley}, \bibinfo{year}{1974}).

\bibitem[{\citenamefont{L{\"u}thi}(2005)}]{luthi05}
\bibinfo{author}{\bibfnamefont{B.}~\bibnamefont{L{\"u}thi}},
  \emph{\bibinfo{title}{Physical Acoustics in the Solid State}}
  (\bibinfo{publisher}{Springer}, \bibinfo{year}{2005}).

\bibitem[{\citenamefont{Khan and Allen}(1984)}]{khan84}
\bibinfo{author}{\bibfnamefont{F.~S.} \bibnamefont{Khan}} \bibnamefont{and}
  \bibinfo{author}{\bibfnamefont{P.~B.} \bibnamefont{Allen}},
  \bibinfo{journal}{Phys. Rev. B} \textbf{\bibinfo{volume}{29}},
  \bibinfo{pages}{3341} (\bibinfo{year}{1984}).

\bibitem[{\citenamefont{Sundaram and Niu}(1999)}]{niu-sundaram-99}
\bibinfo{author}{\bibfnamefont{G.}~\bibnamefont{Sundaram}} \bibnamefont{and}
  \bibinfo{author}{\bibfnamefont{Q.}~\bibnamefont{Niu}},
  \bibinfo{journal}{Phys. Rev. B} \textbf{\bibinfo{volume}{59}},
  \bibinfo{pages}{14915} (\bibinfo{year}{1999}).

\bibitem[{\citenamefont{{Gaidharzhy {\it et al.}}}(2005)}]{mohanty-05}
\bibinfo{author}{\bibfnamefont{A.}~\bibnamefont{{Gaidharzhy {\it et al.}}}},
  \bibinfo{journal}{Appl. Phys. Lett.} \textbf{\bibinfo{volume}{86}},
  \bibinfo{pages}{254103} (\bibinfo{year}{2005}).

\bibitem[{\citenamefont{Mohanty et~al.}(2004)\citenamefont{Mohanty,
  Zolfagharkhani, Kettermann, and Fulde}}]{mohanty-fulde-04}
\bibinfo{author}{\bibfnamefont{P.}~\bibnamefont{Mohanty}},
  \bibinfo{author}{\bibfnamefont{G.}~\bibnamefont{Zolfagharkhani}},
  \bibinfo{author}{\bibfnamefont{S.}~\bibnamefont{Kettermann}},
  \bibnamefont{and} \bibinfo{author}{\bibfnamefont{P.}~\bibnamefont{Fulde}},
  \bibinfo{journal}{Phys. Rev. B} \textbf{\bibinfo{volume}{70}},
  \bibinfo{pages}{195301} (\bibinfo{year}{2004}).

\end{thebibliography}
%

%%%%%%%%%%%%%%%%%%%%%
\end{document}